\documentstyle[prabib,aps]{revtex}
\draft
\begin{document}
\title {Quantization of strongly interacting phonons.}
\author{V.D.Dzhunushaliev
\thanks{E-mail address: dzhun@freenet.bishkek.su}}
\date{}
\address {Theoretical Physics Department \\
          Kyrgyz State National University, Bishkek, 720024}
\maketitle
\begin{abstract}
The assumption is considered that the strong interaction
between phonons makes a certain contribution to the formation of
Cooper pairs. Heisenberg's old idea about the quantization of 
strong nonlinear fields using the Tamm-Dankoff method is discussed. 
The approximate solution method of infinite Tamm-Dankoff
equations system is suggested. This allows us to obtain an equation
for the fixed deformation of the lattice between two Cooper
electrons. Such deformations can introduce a
significant contribution to the energy of Cooper pairs. 
The possible approximate model of the appearance of the flux tube 
(nonlocal object) in the SU(2) Yang-Mills theory is considered. 
The similar mechanism can play important role in High-$T_c$ 
superconductivity if there is the strong nonlinear 
potential.
\end{abstract}

\section{Introduction}
In Ref.\cite{dzh1} a string model of High-$T_c$ superconductivity
was suggested. This model is based on the proposal
that phonons have a strong interaction among themselves.
In this case a tube filled with phonons appears between
Cooper electrons. This is similar to the flux tube filled with
gluons between quarks in quantum chromodynamic. Thus, this
model is based on the existence of strong interaction between
the quantum - carrier of the interaction. The possibility of such
an interaction between phonons in superconductors
is discussed in Ref.\cite{kri} . Such a strong interaction can
obstruct the application of Feynman diagram techniques in this
case. Some time ago W. Heisenberg had conceived of the difficulties
in applying an expansion in small parameters to quantum
field theories having strong interactions. He had been investigated
the Dirac equation with nonlinear terms (Heisenberg equation)
(see, for example, Ref's \cite{heis1} - \cite{heis2}).
In these papers he repeatedly underscored that a nonlinear
theory with a large parameter requires the introduction of another
quantization rule. He worked out a quantization method for
strong nonlinear field unusing the expansion in a small parameter
(Tamm - Dankoff method). It is possible that in High - $T_c$ 
superconductivity the interaction between phonons is strong making
it necessary take into account the interaction between
phonons to correctly calculate the energy of the Cooper pairs.

\section{Heisenberg quantization of field with strong interaction}
\label{a}

Heisenberg's basic idea proceeds from the fact that the n-point Green
functions must be found from some infinity differential equations system
derived from the field equation for the field operator. For example,
we present Heisenberg quantization for nonlinear spinor field.
\par
The basic equation (Heisenberg equation) has the following
form:
\begin{equation}
\gamma ^\mu \partial _\mu  \psi (x) -
l^2 \Im \left [ \psi (\bar  \psi  \psi)\right ] = 0 ,
\label{a1}
\end{equation}
where $\gamma ^\mu $ are Dirac matrices; $ \psi (x)$ is the field
operator; $\bar\psi$ is the Dirac adjoint spinor;
$\Im [ \psi (\bar  \psi  \psi ]=  \psi (\bar  \psi
 \psi)$ or $ \psi \gamma ^5(\bar  \psi \gamma ^5 \psi)$ or
$ \psi \gamma ^\mu (\bar  \psi \gamma _\mu  \psi)$ or
$ \psi \gamma ^\mu \gamma ^5(\bar  \psi \gamma _\mu \gamma ^5
\psi)$. Heisenberg emphasizes that the 2-point Green function
$G_2(x_2,x_1)$ in this theory differs strongly  from the propagator
in linear theory. This difference lies in its behaviour on
the light cone.
$G_2(x_2,x_1)$ oscillates strongly on the light cone in contrast
to the propagator of the linear theory which has a $\delta$-like
singularity. Heisenberg then introduces the $\tau$ functions:
\begin{equation}
\tau (x_1x_2\ldots |y_1y_2\ldots ) =
<0|T\psi (x_1)\psi (x_2)\ldots \psi ^*(y_1)\psi ^*(y_2) \ldots |
\Phi >,
\label{a2}
\end{equation}
where $T$ is the time ordering operator. $|\Phi >$ is a system
state characterized by the fundamental Eq. (\ref{a1}). Relationship
(\ref{a2}) allows us to establish a one - to - one
correspondence between the system state $|\Phi >$ and
the function set $\tau$. This state can be defined using the infinite
function set of (\ref{a2}). Applying Heisenberg's equation (\ref{a1}) to
(\ref{a2}) we can obtain the following infinite equations
system:
\begin{eqnarray}
l^{-2} \gamma ^\mu _{(r)}\frac {\partial}{\partial x^\mu _{(r)}}
\tau (x_1\ldots x_n|y_1\ldots y_n) =
\Im \left [\tau (x_1\ldots x_n x_r|y_1\ldots y_n y_r)\right ] +
\nonumber \\
\delta (x_r - y_1) \tau \left (x_1\ldots x_{r-1}x_{r+1}\ldots x_n|
y_2\ldots y_{r-1}y_{r+1}\ldots y_n\right ) +
\nonumber \\
\delta (x_r - y_2) \tau \left (x_1\ldots x_{r-1}x_{r+1}\ldots x_n|
y_1y_2\ldots y_{r-1}y_{r+1}\ldots y_n\right ) + \ldots .
\label{a3}
\end{eqnarray}
Heisenberg then employs the Tamm - Dankoff method for getting approximate
solutions to the infinite equations system of (\ref{a3}). The key to this
method lies in the fact that the system of equation has an approximate
solution derived after cutting off the infinite equation system
(\ref{a3}) to a finite equation system.
\par
It is necessary to note that a method of solution to Eq. (\ref{a3})
is not important for us. For example, we can try to determine
the Green functions using the numerical lattice calculations.
Here the important point is the following: The technique of expansion
in small parameters (Feynman diagrams) can not be employed
for strong nonlinear fields. It is possible that as in quantum
chromodynamic, where quarks are thought to interact strongly 
by means of flux tubes, so too
in High-$T_c$ superconductivity phonons may strongly interact
among themselves.
\par
In Heisenberg's theory the matter and the interacting fields
are identical: fundamental spinor field $\psi (x)$. From a more
recent perspective this is not the case. An interaction is
carried by some kind of boson field. In superconductivity this
is phonons, in quantum chromodynamic it is the nonabelian $SU(3)$ 
gauge field -- the gluons.
\par
In conclusion of this section we emphasize once more Heisenberg's
statement that the perturbation theory, possibly, is inapplicable
to strong nonlinear field.

\section {Heisenberg quantization method for High-$T_c$
superconductivity}

Thus, the basic assumption advanced here is 
the following: {\sl The energy of Cooper pair has an essential
contribution coming from an interaction of phonons}. This
means that the corresponding sound wave is a nonlinear wave 
(possibly a soliton - like wave). This nonlinear wave is not
a wave between Cooper electrons but wave moving together
with Cooper pairs.
\par
The Lagrangian for phonons (in continuous limit)
can be written down as:
\begin{equation}
L = \int d^3 \vec r
\left [\frac{1}{2}\dot \varphi (\vec r) \dot \varphi (\vec r) -
\frac{c^2}{2} \nabla \varphi (\vec r) \nabla \varphi (\vec r) -
V(\varphi )\right ],
\label{d1}
\end{equation}
here we omit all indexes; $c$ is sound speed;
in an anisotropic superconductor
$\int d^3 \vec r \to \sum \limits _{on \; layer} \int d^2 \vec r$.
In the simplest case we can take
$V(\varphi )= - \lambda \left (\varphi ^2 -
\varphi ^2_0 \right )^2/4$. This can signify that the fixed
deformation of lattice is present between Cooper electrons.
The possibility must not be ruled out that on the classical
level this deformation does not exist but it appears only on the
quantum level (as possibly happens in quantum chromodynamics).
\par
Thus, in this model it is assumed that operators of strong
nonlinear sound waves must satisfy the following
equation (which is implied from the Lagrangian (\ref{d1})):
\begin{equation}
\Box \hat \varphi  = \hat \varphi
\left (\hat \varphi ^2- \varphi ^2_0 \right ),
\label{d2}
\end{equation}
The multitime formalism of Heisenberg's method (when in
$\tau (t_1, t_2, \cdots ), t_1 \ne t_2 \ne \cdots $)
allows us investigate the scattering processes in quantum
theory. The simultaneous formalism (when
$\tau (t_1, t_2, \cdots ), t_1 = t_2 = \cdots =t$)
allows us to calculate the mean value
of the field, the energy, or any combination of field powers.
It is easy to see that the mean value
$<\varphi > = <0|\hat \varphi (x) |0>$
satisfies the following equation:
\begin{equation}
\Box <\varphi (x)> = <\varphi ^3(x)> -
\varphi _0^2<\varphi (x)>.
\label{d3}
\end{equation}
For the definition of $<\varphi ^3(x)>$ we turn to Eq.(\ref{d2}) 
and obtain ($<\varphi ^3(x)> = \tau (xxx$) in Heisenberg's notation):
\begin{equation}
\Box <\varphi ^3(x)> = 3\left (
<\varphi ^5(x)> - \varphi _0^2<\varphi ^3(x)>
\right ),
\label{d4}
\end{equation}
here $<\varphi ^5(x)> = \tau (xxxxx)$.
Analogously it can be used to derive the infinite equation system for
calculating $<\varphi ^n(x)>$. To first approximation we can solve
this equation system using the following  assumption:
\begin{equation}
<\varphi ^3(x)> \approx <\varphi (x)>^3,
\label{d5}
\end{equation}
then we can derive the equation investigated in Ref.\cite{dzh1}.
But now we can take another interpretation of the function,
$\psi (x) = <\varphi (x)>$, as a fixed deformation of lattice.
\par
It should be pointed out that the investigation of
$\tau (xx\cdots ) = <\varphi ^n(x)>$
gives us information about the mean value of the field $\varphi (x)$.
For an investigation of questions on the scattering or interaction
of phonons it is necessary to explore the functions,
$\tau (x_1x_2\cdots ) = <0|\varphi (x_1)\cdots \varphi (x_n)|0>$. 
In following section we discuss the possible mechanism 
of the nonlocal object appearance in the strong nonlinear 
theory on example of SU(2) nonabelian Yang-Mills theory. 
If such a mechanism exist then it can act in the 
High-$T_c$ superconductivity if there we have the strong 
nonlinear potential interaction between phonons. 

\section{Possible mechanism of the tube rise in SU(2) 
Yang - Mills theory}

In this section we propose a model of the tube origin 
in such nonlinear theory as nonabelian Yang - Mills theory. 
At first we consider classical cylindrically symmetric 
solution in this theory \cite{dzh2}. The Yang - Mills 
equations have the following view: 
\begin{equation} 
\nabla _\mu F^{\mu\nu} = 0
\label{5-1} 
\end{equation}
here $\nabla _\mu = \partial _\mu + ig[A_\mu ,F_{\mu\nu}]$ is 
covariant derivative; 
$F_{\mu\nu} = \partial _\mu A_\nu - \partial _\nu A_mu + 
ig[A_\mu , A_\nu]$ is the tensor of Yang - Mills field; 
$A_\mu$ is SU(2) gauge potential. 
\par 
Ansatz for our goals we have as: 
\begin{eqnarray}
A^1_t & = & f(\rho ),
\label{5-2-1}\\
A^2_z & = & v(\rho ),
\label{5-2-2}\\
A^3_{\varphi} & = & \rho w(\rho ),
\label{5-2-3}
\end{eqnarray}
here $\rho ,z, \varphi$ are cylindrical coordinate system. 
After substitution into Eq. (\ref{5-1}) we have the 
following equation: 
\begin{eqnarray}
f'' + \frac{f'}{\rho} & = & fv^2,
\label{5-3-1}\\
v'' + \frac{v'}{\rho} & = & -vf^2,
\label{5-3-2}
\end{eqnarray}
here for simplicity we put $w = 0$. The solutions of these 
classical equations have been investigated in Ref. \cite{dzh2}. 
The asymptotical behaviour of these solutions is following: 
\begin{eqnarray}
f & \approx & 2\left [x + \frac{\cos \left (2x^2 + 2\phi _1\right )}
{16x^3} \right ] ,
\label{5-4-1}\\
v & \approx & \sqrt{2} \frac{\sin \left (x^2 + \phi _1 \right )}{x} ,
\label{5-4-2}
\end{eqnarray}
here $\phi _{1,2}$ are some constants, $x=\rho /\rho _0$ 
is dimensionless radius. 
We see that $f(\rho)$ is increases as $r$ and 
$v(\rho)$ is very strong oscillating function. According to 
Heisenberg ideas of quantization of strong nonlinear theory 
(Heisenberg equation and maybe Yang - Mills theory) we have 
to write Yang-Mills equations as equation for $\hat A_\mu$ 
operator. In order that make the quantization of this 
cylindrically symmetric object we try to shorten the number 
of degrees of freedom which is necessary for approximate 
description of such object to a minimum in the following 
manner:  
\begin{enumerate}
\item 
The degrees of freedom (\ref{5-2-1})-(\ref{5-2-2}) 
contribute significantly to the tube formation. 
\item 
According to (\ref{5-4-1})-(\ref{5-4-2}) we suppose 
that asymptotically $f(\rho)$ function is almost 
classical degrees of freedom but $v(\rho)$ function is 
quantum degrees of freedom. 
\end{enumerate} 
According to Heisenberg ideas the Yang-Mills equations 
for $\hat f(\rho)$ and $\hat v(\rho)$ operators have the 
following view: 
\begin{eqnarray}
\hat f'' + \frac{\hat f'}{x} & = & \hat f \hat v^2,
\label{5-5-1}\\
\hat v'' + \frac{\hat v'}{x} & = & -\hat v \hat f^2,
\label{5-5-2} 
\end{eqnarray}
here $(')$ is derivative with respect to $x$ and. 
Taking into account assumption (2) we have 
\begin{eqnarray}
f'' + \frac{f'}{x} & = & f <v^2>,
\label{5-6-1}\\
\hat v'' + \frac{\hat v'}{x} & = & -\hat v f^2,
\label{5-6-2}
\end{eqnarray}
For receiving the equation for $<v^2>$ we act on 
$\hat v^2(r)$ by operator: 
$(\frac{d^2}{dx^2} + \frac{1}{x}\frac{d}{dx})$: 
\begin{equation} 
(\hat v^2)'' + \frac{1}{x} (\hat v^2)' = 
-2\hat v^2 f^2 + 2 \hat v')^2 
\label{5-7} 
\end{equation} 
Averaging (\ref{5-7}) we have equation for $<v^2>$: 
\begin{equation} 
<v^2>'' + \frac{1}{x} <v^2>' = -2<v^2>f^2 + 2<{v'}^2> 
\label{5-8} 
\end{equation} 
This equation is nonclosed: we have to have once more 
equation for defining $<{v'}^2>$. We suppose that 
$<{v'}^2> \approx \alpha <v^2>$ where $\alpha$ 
is some constant. In this case we have the following 
closed equation set: 
\begin{eqnarray} 
<v^2>'' + \frac{1}{x}<v^2>' = 2<v^2> \left (1 - f^2\right ), 
\label{5-9-1}\\
f'' +\frac{1}{x} f' = f<v^2> 
\label{5-9-2} 
\end{eqnarray} 
here following renaming is made: 
$\alpha ^2 x^2 \to x^2$, 
$<v^2>/\alpha \to <v^2>$, 
$f/\alpha \to f$. 
Only asymptotical solution of these equations is interesting 
for us. It is easy to see that by $x\to \infty$ we 
have: 
\begin{eqnarray} 
<v^2> \approx v_0^2 \frac{\exp (-\gamma x)}{\sqrt{x}}, 
\label{5-10-1}\\ 
f \approx f_\infty + f_0 \frac{\exp (-\gamma x)}{\sqrt{x}}, 
\label{5-10-2}\\ 
f_0 = \frac{f_\infty v_0^2}{2(1 - f_\infty ^2)}, 
\nonumber 
\qquad \gamma = \sqrt{2\left (1 - f_\infty ^2\right )}. 
\end{eqnarray} 
Thus we see that the strong nonlinearity can a cause of 
the appearance of nonlocal object as tube and so on. 
On the basis of this approximate model we can suppose 
that the strong interaction between phonons 
in High $T_c$ superconductivity can lead to the nonlocal 
object appearance, like to the flux tube in quantum chromodynamic, too.

\end{document}